\documentclass[aps,prl,twocolumn,showpacs]{revtex4}
\usepackage[dvips]{graphicx}

\begin{document}

\title{Asymmetric spreading in highly advective, disordered environments.}

\author{John H.\ Carpenter}
\affiliation{
Sandia National Laboratories,
Albuquerque, NM 87185.
}
\author{Karin A.\ Dahmen}
\affiliation{
University of Illinois at Urbana-Champaign,
Department of Physics,
1110 W. Green St, 
Urbana, IL 61801.
}
\affiliation{
Institute for Genomic Biology,
34 Animal Sciences Lab,
1207 W. Gregory Dr.,
Urbana, IL 61801.
}
\date{\today}

\begin{abstract}

Spreading of bacteria in a highly advective, disordered environment is
examined. Predictions of super-diffusive spreading for a simplified
reaction-diffusion equation are tested. Concentration profiles display
anomalous growth and super-diffusive spreading. A perturbation
analysis yields a crossover time between diffusive and super-diffusive
behavior. The time's dependence on the convection velocity and
disorder is tested. Like the simplified equation, the full linear
reaction-diffusion equation displays super-diffusive spreading
perpendicular to the convection. However, for mean positive growth
rates the full nonlinear reaction-diffusion equation produces symmetric
spreading with a Fisher wavefront, whereas net negative growth rates
cause an asymmetry, with a slower wavefront velocity perpendicular to
the convection.

\end{abstract}


\maketitle


%
%

The study of population growth is an integral part of the biological
sciences. Recently the behavior of microbial species, such as
bacteria, has enjoyed much mathematical analysis due to the formation
of intricate equilibrium patterns \cite{Mu93,KoMe96}. Due to environmental and
health concerns the behaviors in driven systems is also of interest.
In this regime the dynamics are typically described using
reaction-diffusion equations which may involve a number of species and
their interactions \cite{Mu93}. While much work has been done on these
types of systems, the inclusion of disorder in the environment has
received limited attention.  This disorder may manifest itself in a
variety of ways, from spatial variations in the available food or in
the presence of poisons to random diffusion constants.

In particular, a reaction diffusion equation with spatially
varying growth factors may take the following form,
\begin{eqnarray}
\label{eq:rd}
\partial_t c(\vec{x},t) &=& D\nabla^2c(\vec{x},t) - \vec{v} \cdot \nabla c(\vec{x},t) \nonumber \\
& & +[a+U(\vec{x})]c(\vec{x},t)-bc^2(\vec{x},t),
\end{eqnarray}
where the $U(\vec{x})$ are spatially random growth rates and the
convection velocity $v$ drives the population through the environment
\cite{NeSh98}. If both $U=0$ and $v=0$ then Eq.\ \ref{eq:rd} reduces
to the Fisher equation, where the growth rate $a$ provides exponential
growth which is cut off by the nonlinear interaction term $b$ at the
system's carrying capacity, $a/b$ \cite{Mu93}.  The linear regime of
Eq.\ \ref{eq:rd}, around the fixed points $c=0$ and $c=a/b$, has been
studied in detail in Ref.\ \cite{NeSh98}. Only some limited numerical
simulation have been performed for the nonlinear case
\cite{DaNeSh99}. In the limit of large convection velocity $v$, fixed in the $y$ direction, and with
$b=0$, a substitution of the form
\begin{equation}
\label{eq:sub}
c(\vec{x},y,t) = \frac{1}{\sqrt{4 \pi Dt}}exp \left (at-\frac{(y-vt)^2}{4Dt} \right) W(\vec{x},y),
\end{equation}
allows one to obtain a simplified form of Eq.\ \ref{eq:rd},
\begin{equation}
\label{eq:simp}
v \partial_t W(\vec{x},t) = D\nabla^2 W(\vec{x},t) + U(\vec{x},t)W(\vec{x},t).
\end{equation}
Here $y$ is relabeled as $t$ and the directions perpendicular to $y$ as
$\vec{x}$ \cite{NeSh98}. As the substitution (Eq.\ \ref{eq:sub})
contains the exponential growth and diffusion in the $y$ direction,
the function $W(\vec{x},t)$ describes the cross section perpendicular
to the convection for a population at the 'time' $t=y/v$.
Interestingly, this simplified equation has the form of an imaginary
time Schr\"odinger equation with a random, fluctuating potential.
Additionally it is directly connected to the problem of directed
polymers in random media \cite{HeZh95}. It has been shown that Eq.\ 
\ref{eq:simp} predicts super-diffusive growth for the long time, large
distance behavior, with an exponent of $2/3$ in one spatial dimension
\cite{NeSh98}. This exponent has been reproduced numerically by
examining the averaged mean squared displacement of the optimal path
(lowest energy path) in directed polymers \cite{HeZh95}.  Through a
detailed examination of the full concentration profiles of Eq.\ 
\ref{eq:simp}, this report examines the behavior of this simplified
equation in the context of population growth with the goal of
obtaining a better understanding of the full equation's (Eq.\ 
\ref{eq:rd}) behavior in both the linear ($b=0$) and nonlinear
($b>0$) regimes.  First a perturbation analysis of Eq.\ 
\ref{eq:simp} yields a crossover time dividing pure diffusion and the
super diffusive behavior. Afterwards, one dimensional numerical
simulations describe concentration profiles and anomalous growth of
Eq.\ \ref{eq:simp} and test the predictions for the diffusion exponent
and crossover time.

%
%

To obtain a perturbation expansion for Eq.\ \ref{eq:simp}, first note
that it is an initial value problem. Thus, in the spirit of Ref.\
\cite{FoNeSt77}, one applies a Fourier-Laplace transform, \begin{equation}
\label{eq:tran}
\widehat{W}(\vec{k},\omega) = \int_0^{\infty}dt e^{-\omega t} \int_{-\infty}^{\infty}d^dx e^{-i\vec{k}\cdot\vec{x}}W(\vec{x},t).
\end{equation}
Equation \ref{eq:simp} then takes the form,
\begin{eqnarray}
\label{eq:fltsimple}
\lefteqn{\widehat{W}(k,\omega)  =  v G_0(k,\omega)\widetilde{W}(k,0)} \nonumber\\
&&+\ G_0(k,\omega) \!\int\! d^dq \!\int\! dt \!\int\! d\Omega_a
\!\int\! d\Omega_b\ e^{-(\omega-(\Omega_a+\Omega_b)) t} \nonumber \\
&&\qquad \times \widehat{U}(q,\Omega_a) \widehat{W}(q,\Omega_b),
\end{eqnarray}
with the abbreviations $\int d^dq \equiv \int_{-\infty}^\infty
\frac{d^dq}{(2\pi)^d}$, $\int dt \equiv \int_0^\infty dt$, and $\int
d\Omega_y \equiv \int_{y-i\infty}^{y+i\infty}\frac{d\Omega_y}{2\pi i}$
where $y = a$ or $b$ where $G_0(k,\omega) = (v\omega+Dk^2)^{-1}$ is
the free propagator and $\widetilde{W}$ denotes taking only the
Fourier transform. In obtaining this form, the Bromwich integral
giving the inverse Laplace transform was used.

\begin{figure}
\begin{tabular}{l c}
\includegraphics[angle=0,width=1.0cm]{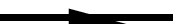} $\quad = \quad$
\includegraphics[angle=0,width=1.0cm]{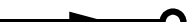} $\quad + \quad$
\includegraphics[angle=0,width=2.0cm]{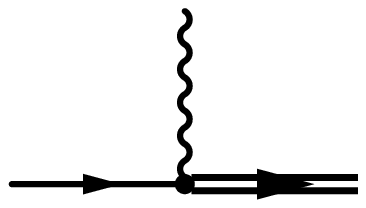} $\quad$ 
& (a) \\
&\\
&\\
$\overline{
\includegraphics[angle=0,width=1.0cm]{simpeqgrapha1.eps}} \quad = \quad$
\includegraphics[angle=0,width=1.0cm]{simpeqgrapha2.eps} $\quad + \quad$
\includegraphics[angle=0,width=3.0cm]{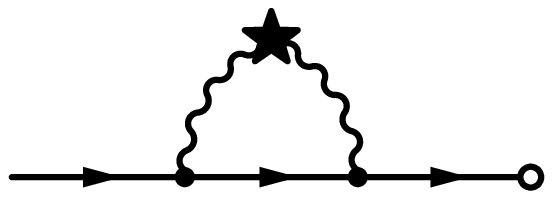} $\quad$ 
& (b)
\end{tabular}

\caption[Diagrammatic representation of simplified equation]{
\label{fig:simpeqgraph}
Diagrammatic representation of the Fourier-Laplace transformed,
simplified equation (Eq.\ \ref{eq:fltsimple}). Part (a) gives the
exact equation while part (b) displays the disorder averaged
perturbation series to one loop order.  
}
\end{figure}

A graphical representation of Eq.\ \ref{eq:fltsimple} and its second
order, disorder averaged expansion are shown in Fig.\
\ref{fig:simpeqgraph}(a) and (b) respectively.
As $U$ has zero mean, upon taking the disorder average the first order
term drops out. Hence determining to one loop order the renormalized
propagator $G_R$, defined as $\widehat{W}(k,\omega) = v
G_R(k,\omega)\widetilde{W}(k,0)$, requires calculating only the second
order term. For a uniform distribution of width $\Delta$ the
correlator is $\overline{U(\vec{x},t)U(\vec{x}',t')} =
\frac{\Delta^2}{12}l_x^d l_t\delta^d(x-x')\delta(t-t')$ where $l_y$ is the
lattice constant for the $y$ direction. With this fact and some
straightforward contour integrations, the renormalized propagator of
Fig.\ \ref{fig:simpeqgraph}(b) becomes,
\begin{equation}
\label{eq:gr}
G_R(\vec{k},\omega) = G_0(\vec{k},\omega) + \frac{S_d \Delta^2 l_t}{48dv} G_0^2(k,\omega).
\end{equation}
where $S_d$ is the surface area of a $d$-dimensional sphere of unit radius.  Expanding $G_0$
and $G_R$ for $k \rightarrow 0$ yields 
\begin{equation}
\label{eq:dr}
D_R = D\left(1+\frac{S_d \Delta^2 l_t}{24dv^2\omega}\right).
\end{equation}

When the second term on the right side of Eq.\ \ref{eq:dr} becomes on
the order of one, then pure diffusion is no longer the
dominant term. The crossover time is proportional to the value of $\omega^{-1}$ at
this point. From Eq.\ \ref{eq:dr} the crossover time $T$ is given by,
\begin{equation}
\label{eq:crosstime}
T = \frac{48\pi dv^2}{S_d l_t \Delta^2}.
\end{equation}
The crossover time depends on both the velocity of the flow as well as
the width of the distribution of random growth rates. When the
velocity increases, the system is pushed through the random
environment before it has time to experience the fluctuations, seeing
an effectively averaged environment. Hence it makes sense that $T$ is
increased by higher velocities. On the other hand, increasing the
width of the random distribution of growth rates creates optimal paths
in the system which have larger effective growth rates. This causes
pure diffusion to break down earlier, hence the inverse dependence
with the crossover time. 

%
%

Numerical simulations of Eq.\ \ref{eq:simp} were performed in one
spatial dimension using a Runge-Kutta technique \cite{NRC}. A Gaussian
initial condition of unit variance was centered on a lattice of
$20000$ sites. This was large enough to insure that the boundaries
were never encountered by the concentration. The random growth rates
depend on time, and so must be updated during the simulation. To
provide equal time and space lattice constants, a Runge-Kutta step
size of $0.1$ was chosen and the growth rates were updated after every
ten time steps. Lastly, the concentrations were normalized after every
time step.

\begin{figure}
\begin{center}
\includegraphics[angle=0,width=4.1cm]{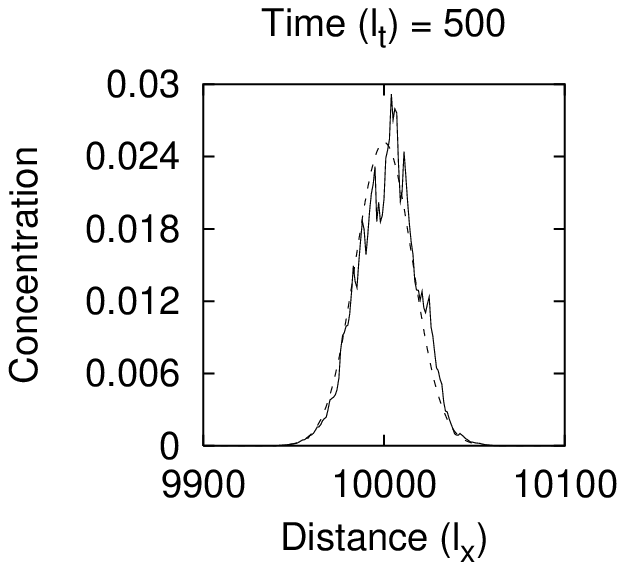}
\includegraphics[angle=0,width=4.1cm]{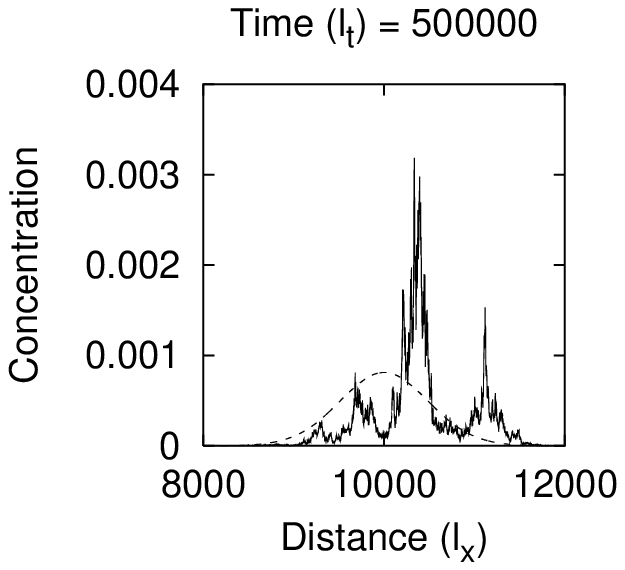}\\
\includegraphics[angle=0,width=4.1cm]{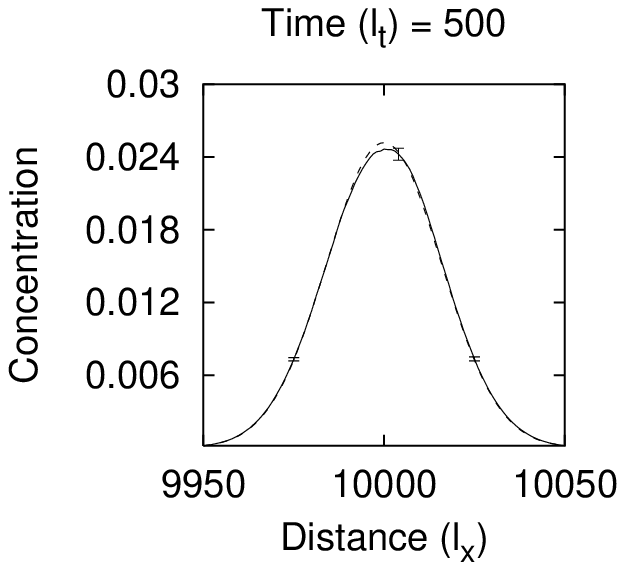}
\includegraphics[angle=0,width=4.1cm]{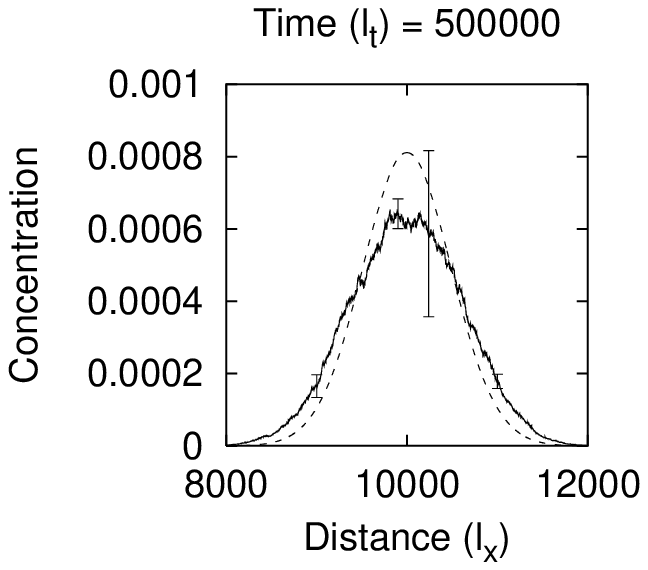}
\end{center}
\caption[Concentration profiles for one dimensional spreading]{
\label{fig:1dprofiles}
Concentration profiles for one dimensional spreading according to Eq.\ 
\ref{eq:simp}, normalized by the total population. The top plots are
for a single disorder realization while the bottom plots are averaged
over $1000$ random configurations. A Gaussian of unit variance was
used as the initial profile.  The solid lines are profiles for a
disordered environment with $v=4.0$ and $\Delta = 1.0$ and the dashed
lines show the purely diffusive case.  A maximal and several
characteristic error bars are shown on the averaged distributions.  
}
\end{figure}

The initial concentration profile mimics the inoculation of a medium
with an initial bacterial sample. With no disorder present the
Gaussian shape would persist, with a variance increasing in
time. However, the disorder destroys this by providing particularly
favorable paths along which growth may occur. In Fig.\
\ref{fig:1dprofiles} concentration profiles are shown for two
different times after inoculation. 
The top two plots contain profiles for a single disorder
realization. For the short time there is very little deviation from
the purely diffusive situation. On the other hand, the large time
concentration profile deviates greatly, with large spikes developing. 
These regions correspond to the end of a path in the $(x,t)$
space which had particularly favorable growth rates and thus resulted
in a much larger population of bacteria than would be expected from a
homogeneous environment. For the long times, these paths may end with
increasing distance from the original starting position effectively
shifting the mean position of the concentration from the starting
position. There may also be several competing paths that have nearly
the same effective growth rate resulting in several concentration
spikes in the profile. These behaviors do not occur in the homogeneous
case as the spreading Gaussian profile always remains centered on the
starting point and symmetric about that point. Upon performing a
disorder average, as seen in the bottom of Fig.\ \ref{fig:1dprofiles},
the large, off-center peaks result in average concentration profiles
whose widths increase faster then the purely diffusive case.

The concentration profiles in Fig.\ \ref{fig:1dprofiles} were
normalized to have a clear comparison with the case of pure diffusion.
As no growth terms are present in Eq.\ \ref{eq:simp} when $U=0$, the
latter case remains normalized. This breaks down in the former case
however. Even though $U$ has a zero average, fluctuations in the
growth factors lead to anomalous growth. Although the effective growth
is relatively small, corresponding to $a=0.00223/l_t$ in Eq.\ \ref{eq:rd}
for the system shown in Fig.\ \ref{fig:1dprofiles}, at the longest
times ($t=500000l_t$) the effects are profound as the total population
becomes on the order of $e^{1100}$. Clearly a diverging bacterial
density is unphysical. It emerges here because the nonlinear death
term has been dropped in the simplified equation, Eq.\ \ref{eq:simp}.

A simple argument for the appearance of this anomalous growth lies in
the asymmetry between the growth and death processes. In particular,
consider a small concentration of bacteria present in a favorable
environment, $U>0$. This concentration will grow exponentially in time
and will spread additional concentration to neighboring areas via
diffusion. On the other hand, in an unfavorable environment for
growth, with $U<0$, the local population will experience an
exponential decay in concentration. While this decreases the total
local population, the decrease imposed in neighboring areas is not the
opposite of the growth case. The asymmetry lies in the fact that the
local (and total) population cannot fall below zero. One cannot have a
negative number of organisms. Hence the difference in concentration
between neighboring sites, proportional to the rate of transport, is
smaller with $U<0$ leading to a reduced rate of population loss
compared to the gain in population when $U>0$.  Therefore the
asymmetry of the diffusion process in the random environment leads to
an effectively increased growth rate.

\begin{figure}
\begin{center}
\includegraphics[angle=0,width=8.2cm]{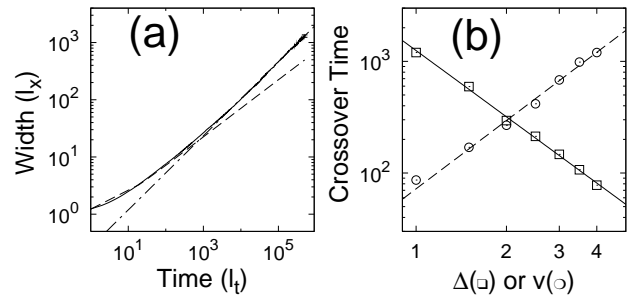}
\end{center}
\caption[One-dimensional super-diffusive behaviors]{
\label{fig:1dwct}
Super-diffusive behaviors for the simplified equation in one dimension:
(a) Concentration width vs. time averaged over $1000$ disorder
configurations with $v=4.0$ and $\Delta = 4.0$ (solid line) and the
purely diffusive case (dashed line). The two curves begin with a
power-law of $0.5$, but for long times the disordered case switches to
a power-law of $0.647 \pm 0.001$, shown as a
dot-dash line. (b) Crossover times for variation of
$\Delta$ with $v=4.0$ (squares) and variation of $v$ with
$\Delta=1.0$ (circles). The solid line shows a power law of $-1.97\pm0.05$ and
the dashed line a power law of $2.03\pm0.10$ as fit to the data
varying $\Delta$ and $v$ respectively. Each point is averaged over
$3000$ random configurations with statistical error bars on the order
of the symbol size.
}
\end{figure}

To measure the diffusion exponent, the width of the disorder averaged
concentration profile, as shown in Fig.\ \ref{fig:1dprofiles},
was measured as a function of time. The resulting curve is shown in
Fig.\ \ref{fig:1dwct}(a).
For pure diffusion the profile width grows as $t^{0.5}$ as one
expects. The disordered case is different, with two regions of clearly
different power-law behaviors.  For small times the disordered width
follows the behavior of the purely diffusive case.  However, as the
time becomes large, the disordered case deviates from pure diffusion
and instead grows with a power-law exponent equal to $0.647 \pm
0.001$. This super-diffusive behavior is in good agreement with the
exponent value $2/3$ that has been previously predicted \cite{NeSh98}. As
explained above, this super-diffusive behavior arises due to the
appearance of optimal growth rates that deviate far from the center of
the population. Upon averaging, these shift concentration from the
center of the profile resulting in a width wider then the diffusive
case.

A clear crossover to super-diffusive behavior is seen in Fig.\ 
\ref{fig:1dwct}(a). However, the location of this crossover depends on
the simulation parameters. For the simulations, the predicted
crossover time, Eq.\ \ref{eq:crosstime}, becomes $T = 24\pi
v^2/\Delta^2$. Compared to the visually apparent crossover point in
Fig.\ \ref{fig:1dwct}(a), the prediction of $T\approx 75$ is roughly an
order of magnitude too small. This should not be completely unexpected
as Eq.\ \ref{eq:crosstime} really describes the time where departure from
purely diffusive behavior begins. At this point the width is growing
super diffusively, but does not saturate at the full exponent until $t
\gg T$.

The crossover time's dependence on the velocity and random width of the
growth rates is shown in Fig.\ \ref{fig:1dwct}(b).
These times were obtained at the point where the difference in width
between the disordered and purely diffusive case was equal to that of
a baseline case ($v=4$ and $\Delta=1$) at roughly the crossover time
predicted by Eq.\ \ref{eq:crosstime}, $t=1200$.  The crossover times
agree very well with the quadratic behavior, $T \sim (v/\Delta)^2$,
predicted by Eq.\ \ref{eq:crosstime}. From Fig.\ \ref{fig:1dwct}(b) the
variation with $\Delta$ at fixed $v$ behaves as $T \sim
\Delta^{-1.97\pm0.05}$ and the variation of $v$ at fixed $\Delta$
results in a power-law $T \sim v^{2.03\pm0.10}$.

\begin{figure}
\begin{center}
\includegraphics[angle=0,width=4.1cm]{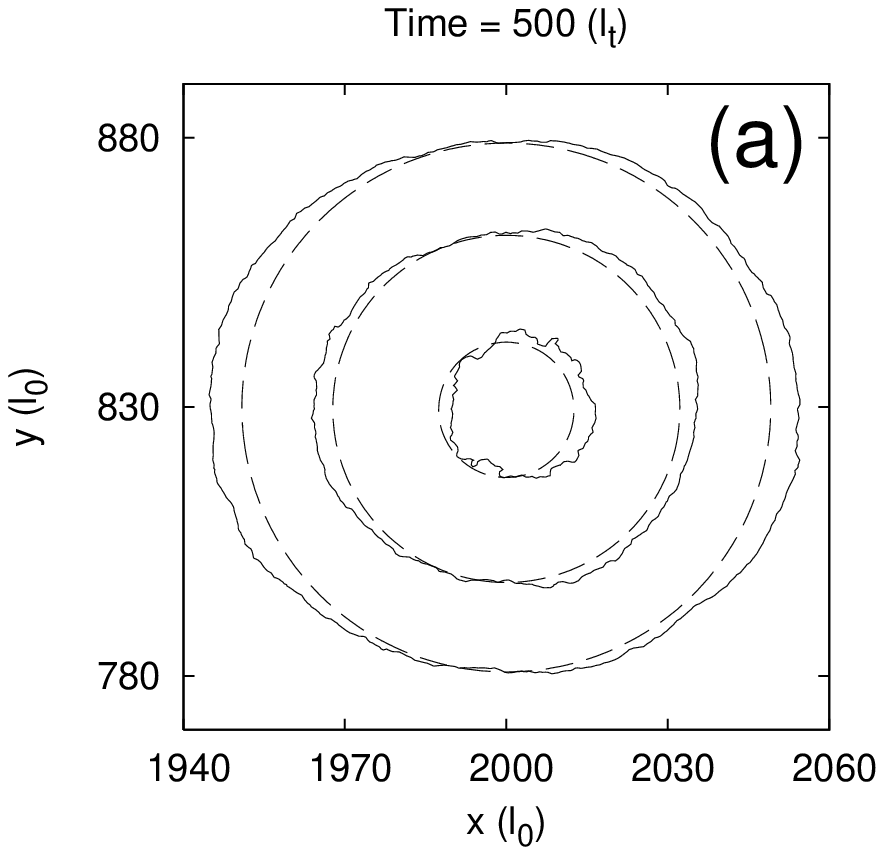}
\includegraphics[angle=0,width=4.1cm]{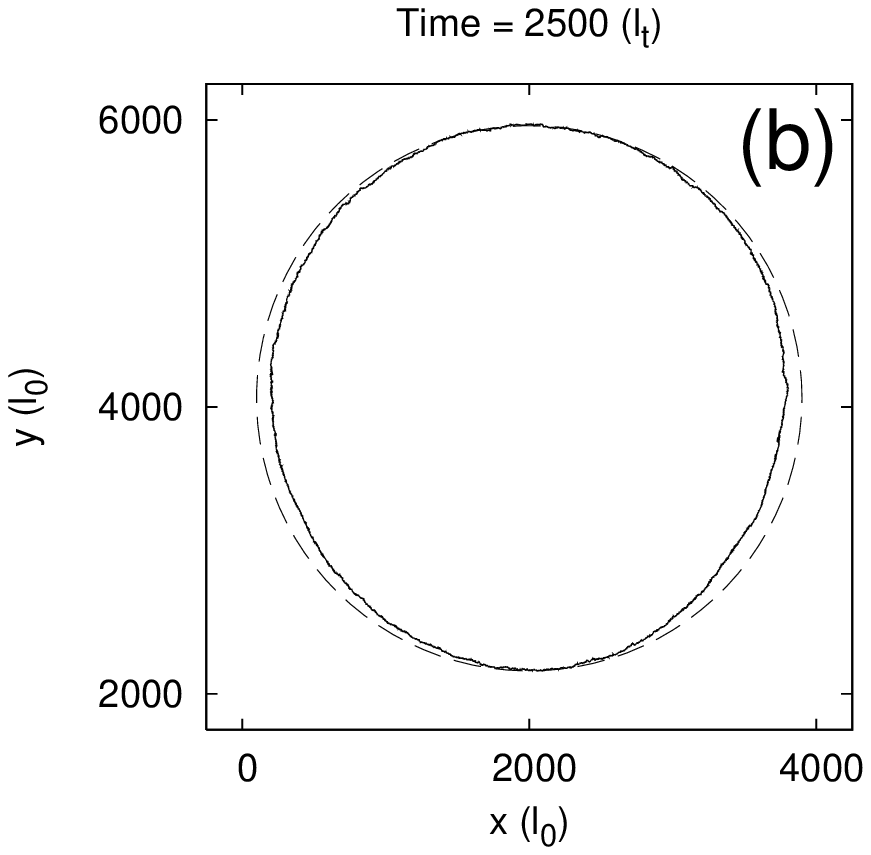}
\end{center}
\caption[Two-dimensional asymmetric spreading]{
\label{fig:2d}
Two-dimensional asymmetric spreading for the full reaction-diffusion
equation (Eq.\ \ref{eq:rd}) with convection along the vertical axis:
(a) disorder averaged concentration maps for the linear case, $b=0$,
show super-diffusive spreading perpendicular to the convection
velocity for $\Delta=2$, whereas (b), the nonlinear case $b>0$ with
$a<0$ and $\Delta=3$, has a wavefront that propagates more slowly in
the direction perpendicular to the convection, resulting in the
opposite asymmetry. For comparison, the dashed lines show the
homogeneous, linear case in (a), and a circle, such as is obtained for
the nonlinear case with $b>0$ and $a>0$, in (b).  }
\end{figure}

As the simplified equation describes the cross section of Eq.\
\ref{eq:rd} perpendicular to the convection, it implies that Eq.\
\ref{eq:rd} with $b=0$ should exhibit super-diffusive behavior in that
direction. Indeed, as seen in Fig.\ \ref{fig:2d}(a), the concentration
contours for a two-dimensional simulation of this linear case shows
contours of width equal to the homogeneous case in the direction
parallel to the convection, but spreading faster perpendicular to
it. 

As noted above, the linear case is not physical for long times due to
unrealistic organism densities. The nonlinear case with $b>0$ presents
a much different spreading picture. For $a>0$ a symmetric Fisher wave
\cite{Mu93} develops. This symmetry should not be completely
unexpected. In the linear case, the mechanism for the enhanced
spreading perpendicular to the convection was the disorder averaging
of the large, asymmetric concentration spikes, such as shown in Fig.\
\ref{fig:1dprofiles}. Since these spikes are cut off by the carrying
capacity, this behavior is absent in the nonlinear case and spreading
is symmetric. The spreading is still enhanced, however, as the
wavefront velocity increases with increasing disorder. In particular,
outside the wavefront the linear regime applies and enhanced growth is
found. Thus, one may argue \cite{cthesis} that the growth rate in the
Fisher velocity expression should be replaced by the real growth rate,
giving the wavefront velocity
\begin{equation}
\label{eq:wf}
v_\mathrm{wf}=2\sqrt{(a_\mathrm{eff}(\Delta)+a)D}.
\end{equation}
Here $a_\mathrm{eff}$ is the effective growth rate of the
corresponding linear problem which depends on the disorder strength
$\Delta$. Numerical simulations of Eq.\ \ref{eq:rd} for a range of
disorders find excellent agreement with this wavefront velocity
\cite{cthesis}.

The wavefront velocity expression, Eq.\ \ref{eq:wf}, has an important
implication. Namely, attempting to poison or destroy a colony of
organisms, by applying $a<0$, may fail if the disorder
creates sufficiently enhanced growth, $a_\mathrm{eff}>-a$. Even more
interesting, the resulting wavefront is asymmetric but in the opposite
manner to the above linear case. Figure \ref{fig:2d}(b) shows the
wavefront obtained from a numerical simulation of Eq.\ \ref{eq:rd} with
$b>0$ but $a<0$. The direction parallel to the convection has a
wavefront velocity that follows Eq.\ \ref{eq:wf}, but the
perpendicular wavefront velocity is smaller, resulting in an
asymmetric droplet. Qualitatively, the smaller wavefront velocity
arises due to the loss of optimal growth paths. The net negative growth
rate does not allow paths passing regions of random, negative growth
rates which were previously possible due to the additional positive
growth factor. The parallel direction is unaffected, because detours
around these lost paths may take place on each side of the lost
paths. However, if the lost path occurs on the edge of the growing
droplet, only paths nearer the droplet center remain, resulting in a
reduced spreading speed.

\begin{acknowledgments}
The authors thank David Nelson and Nadav Schnerb for very useful
discussions.  The work was supported by NSF grant DMRs 03-25939ITR
(MCC), 00-72783, and 03-14279, an A.\ P.\ Sloan fellowship (to K.\
D.), and an equipment award from IBM.
\end{acknowledgments}

\end{document}